\documentclass[12pt,onecolumn,showpacs,amssymb,aps,nofootinbib,floatfix]{revtex4-1}
\usepackage{epsfig}
\usepackage{xcolor}
\usepackage[utf8]{luainputenc}

\begin{document}
\title{Color reconnection effects on resonance production}

\author{R.~Acconcia}
\author{D.D.~Chinellato}
\author{R.~Derradi de Souza}
\affiliation{Universidade Estadual de Campinas, S\~{a}o Paulo, Brazil}

\author{C.~Markert}
\affiliation{The University of Texas at Austin, Austin, Texas, USA}

\author{J.~Takahashi}
\author{G.~Torrieri}
\affiliation{Universidade Estadual de Campinas, S\~{a}o Paulo, Brazil}

\begin{abstract}

We present studies that show how multi-parton interaction and color reconnection
affect the hadro-chemistry in proton-proton (pp) collisions with special focus on the
production of resonances using the PYTHIA8 event generator.
We find that color reconnection suppresses the relative production of meson
resonances such as $\rho_{0}$ and K*, providing an alternative 
explanation for the K*/K decrease 
observed in proton-proton collisions as a function of multiplicity by the ALICE collaboration.
Detailed studies of the underlying mechanism causing meson resonance suppression 
indicate that color reconnection leads to shorter, less energetic strings whose 
fragmentation is less likely to produce more massive hadrons for a given quark content, 
therefore reducing ratios such as K*/K and $\rho_0/\pi$ in high-multiplicity pp collisions. 
In addition, we have also studied the effects of allowing string junctions to form and 
found that these may also contribute to resonance suppression. 


\end{abstract}
\maketitle


\section{Introduction}
 
In heavy-ion experiments, observables are studied as a function of collision centrality, defined by the measured event charged particle multiplicity, to evaluate the dependence with system size and are also compared to elementary proton-proton collisions that is used as reference data, since no effects from Quark-Gluon Plasma is expected in such small systems. 
However, the study of proton-proton (pp) and proton-lead (pA) collisions as a function of event activity, that is also quantified by the event charged particle multiplicity, has recently gathered much attention. 
Features reminiscent of heavy-ion phenomenology, such as collective flow~\cite{smallflow1,smallflow2,smallflow3}, baryon anomaly~\cite{pabaryon} and increase in relative strangeness production~\cite{nature} have shaken the foundation of the earlier picture of ``elementary'' pp probes. 
In particular, the observed relative enhancement of strange baryons as a function of event multiplicity in pp collisions raises the question of to what extent the hadronization and system evolution in smaller systems is similar to that of larger ones~\cite{becattini}. 

One promising observable for the latter question is the study of short-lived hadronic resonances. 
In heavy-ion collisions, such resonances not only convey information regarding hadro-chemistry at hadronization but are also sensitive to the hadronic rescattering phase that takes place prior to final kinetic freezeout~\cite{markert,mebell}. 
Indeed, the observed experimental suppression of resonances as a function of collision centrality has made this interpretation widely accepted~\cite{resorhic,resoalice}, although the lack of a systematic pattern of suppression prevented a definite conclusion to be reached. 
Models that include a microscopic hadronic phase, such as EPOS~\cite{epos}, are able to describe the resonance suppression depending on their decay life-time and the increasing volume size. 
In this case the resonance suppression is only interpreted as an effect of the hadronic phase. 
The question whether the observed resonance suppression is due to a partonic phase is still open.
Furthermore, the observation of resonance suppression in high multiplicity pp and pA events~\cite{resosmall}  has added an additional challenge to the rescattering hypothesis, since even high multiplicity events, due to their smallness in configuration space, are not expected to have the same amount of rescattering than larger AA events~\cite{meridge}. 
Since hadro-chemistry is intimately connected with hadronization, it is difficult to see how the purely partonic initial state models, advocated in the generation of other signatures of collectivity in small systems (for example~\cite{cgc}), could apply in this case. 
In addition, the question if a long lived hadronic system, such as the ones created in heavy-ion collisions, needs a collective partonic interaction phase to describe the data is still open.

One mechanism which is known to affect both hadro-chemistry and cause collectivity-like signals in small systems \cite{mexicans} is color reconnection (CR), long implemented in string fragmentation-based Monte Carlo event generators~\cite{pythia1,pythia2}. 
For energetic collisions, different multi-parton interactions (MPI) evolve into different strings. 
The subsequent inter-connections and interactions of these strings from different MPI have been known to produce some signatures associated with collectivity. 
Color reconnection describes very well the observed rising of the average charged particle $p_{T}$ in pp events with event activity~\cite{Jesper2015A}. 
Other effects such as an increase of event-by-event fluctuations of the mean transverse momentum was also observed as a result of CR~\cite{ptfluct}.
In ~\cite{Jesper2015B}, it is shown that CR also modifies the final string fragmentation into hadrons and in particular, an enhancement of baryons over mesons is observed with increasing multiplicity.
Therefore, resonance measurements call for a study of the effects of MPI and CR on hadro-chemistry, with a special focus on how these mechanisms affect hadronization for different particle properties like mass, spin and strangeness content.

In this work, we use the PYTHIA8 event generator~\cite{pythia1,pythia2} version 8.226 with the Monash 2013 tune to simulate inelastic proton-proton collisions at 7 TeV and study the effects of CR on particle chemistry. 
Each simulated event is required to have produced at least one 
charged particle within $|\eta|<1.0$ to match what is done in recent identified particle measurements from the 
ALICE Collaboration~\cite{nature}. 
We then compare (a) a configuration in which no CR is allowed to take place, (b) the default `MPI-based' CR algorithm and (c) one in which the latest 
CR model, denoted in this work by `More QCD-based CR model' \cite{moreqcd}, is used. 
In the latter CR scheme, hadronizing strings are allowed to reconnect, forming quark junctions that increase baryon production, resulting in a better reproduction of the observed $p_{T}$-differential 
$\Lambda/K^{0}_{S}$ ratio. 
Emphasis will be given on the resonance to non-resonance $p_{T}$-integrated particle ratio K*/K, for which measurements as a function of multiplicity already exist in the public domain.

\section{Effects of color reconnection on the K* resonance}

\begin{figure*}\begin{center}
\includegraphics[width=0.77\textwidth]{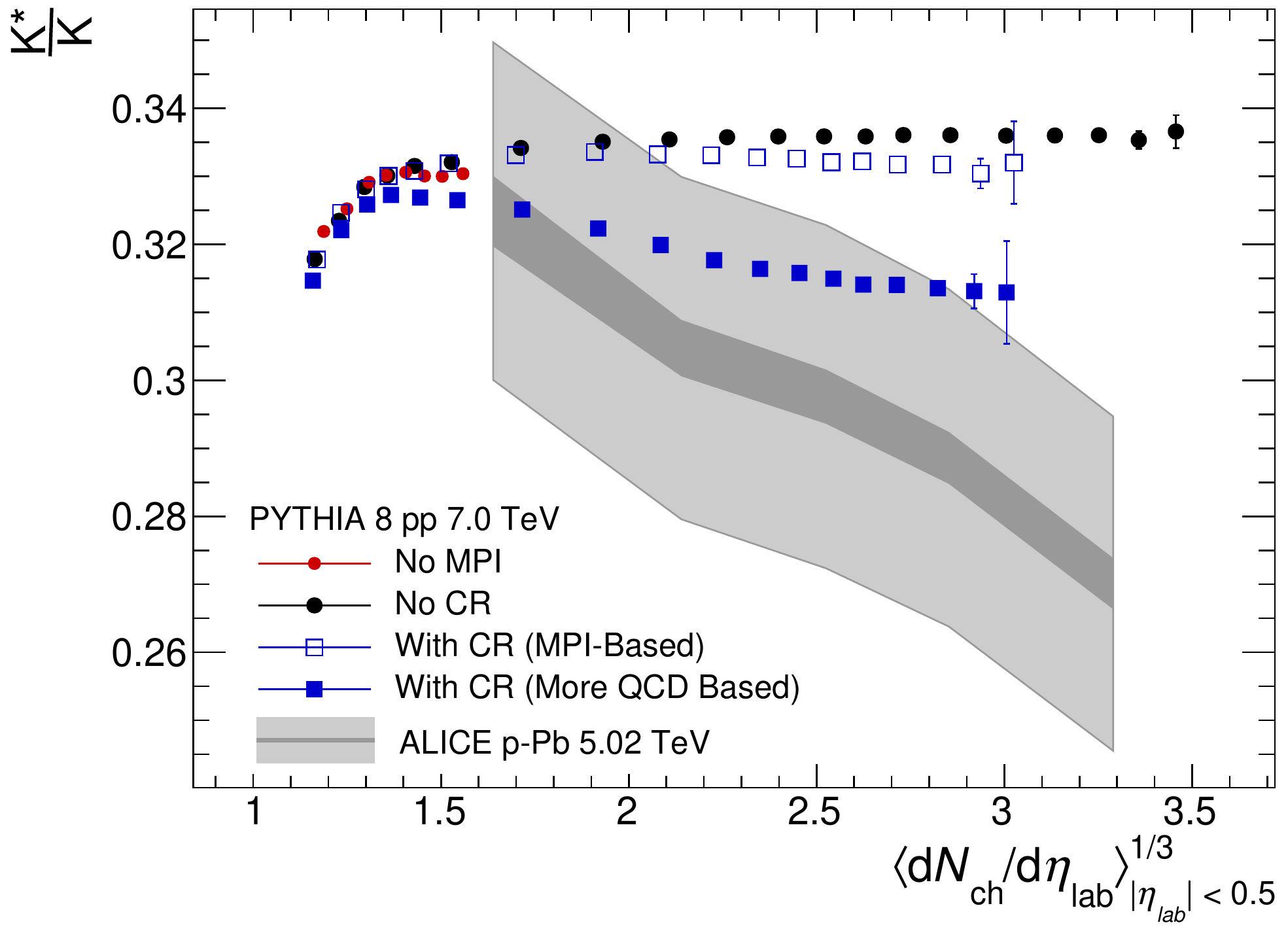}
\caption{\label{KStarOverKVsData}  K*/K $p_{T}$-integrated ratio as a function of charged particle multiplicity as measured by the ALICE experiment in p-Pb collisions at $\sqrt{s_{NN}}=5.02$~TeV \cite{resopPb}  compared to PYTHIA8 predictions with (blue squares) and without (black circles) color reconnection.}
\end{center}\end{figure*}

In Fig.~\ref{KStarOverKVsData}, we present the K*/K integrated ratio as a
function of multiplicity predicted by PYTHIA8 for pp collisions at 7~TeV without
CR (black squares), with `MPI-based' (blue hollow squares) and `More QCD-based' (blue solid squares) compared
to measurements by ALICE (shaded gray area) in p-Pb collisions at
$\sqrt{s_{NN}}=5.02$~TeV \cite{resopPb}. While of course these are two different
colliding systems and energies, a preliminary version of the K*/K ratio in pp
collisions at 7~TeV follows the same trend as the p-Pb data \cite{resosmall}, 
and therefore using the p-Pb data as a proxy for the
as of yet unpublished pp measurement is reasonable. 
For completeness, we also compared results obtained from PYTHIA8 without MPI
(red solid circles). The results obtained without MPI concentrate in the low multiplicity 
region and are consistent with the values
obtained with MPI, regardless of if CR is enabled or not. This indicates that for very low
multiplicities, the dominating physics is that of a single partonic interaction, which is understandably 
different than the one obtained in the MPI regime. In this single interaction limit, variables such as transferred 
momentum may play a more significant role, justifying the increasing trend observed for very low multiplicity 
in Fig.~\ref{KStarOverKVsData} which runs counter to the decreasing trend observed for larger multiplicities, where
one expects a stronger influence from MPI and CR in measured particle yields. 

An important consideration when comparing to experimental data is the counting of particles that are decay products. 
In order to generate PYTHIA8 predictions that are comparable to measurement, we allowed decays only for species for which $c\tau<3mm$, therefore including weak decays from heavy flavor but excluding strangeness decay products.

The K*/K ratio in real data decreases by as much as $\sim$15\% with increasing multiplicity, which is not fully matched by PYTHIA8 predictions in any of the configurations tested here. 
However, it has to be noted that the addition of CR reduces the ratio at high multiplicity by up to 7\% for the `More QCD-based' scheme and remarkably also introduces a downward trend in 
the K*/K with multiplicity. This decrease is also present when using `MPI-based' CR, albeit smaller in magnitude, suggesting that there may be a mechanism for suppressing this ratio 
that, while quantitatively different in the two cases, may always be present. 
This is an important observation from the modeling perspective, since this mechanism is also able to suppress relative K* production and is fundamentally different from the rescattering 
explanation usually found in literature. 

\begin{figure*}\begin{center}
\includegraphics[width=0.7\textwidth]{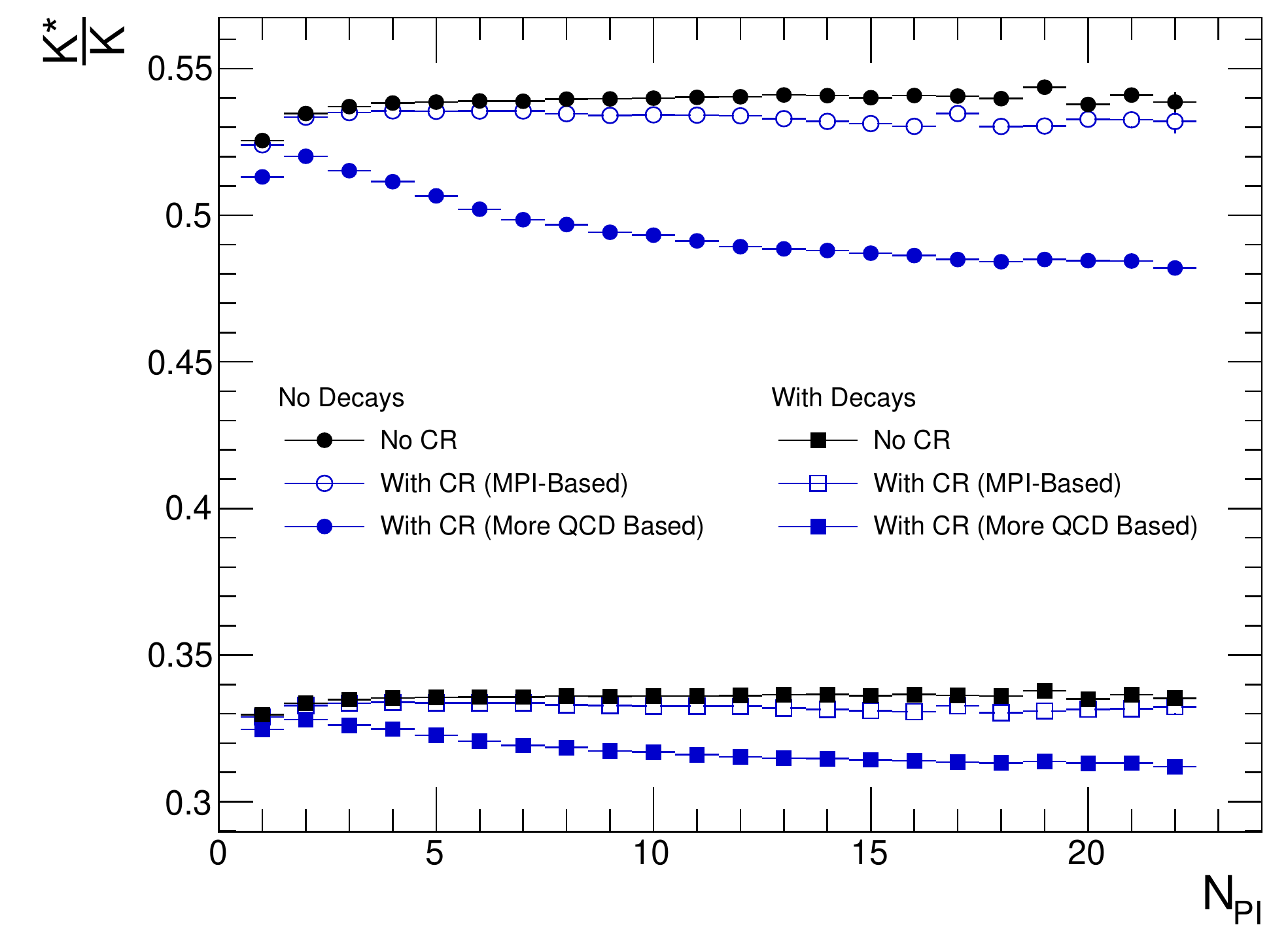}
\caption{\label{decayissue}   K*/K $p_{T}$-integrated ratio as a function of number of partonic interactions $N_{PI}$, for pp collisions at 7 TeV obtained with PYTHIA8 with (blue symbols) and without (black symbols) color reconnection, and with (squares) and without (circles) considering hadronic decays.}
\end{center}\end{figure*}

In order to isolate the effect of CR on the K*/K ratio, we study the dependence on the number of partonic interactions ($N_{PI}$) as shown in Fig.~\ref{decayissue}, represented by the square symbols, where the black symbols represents the ratio for no CR, and the blue symbols represent the ratio for when the various CR algorithms are turned on. 
The evolution of the ratio with $N_{PI}$ is similar to its evolution with charged particle density seen in Fig.~\ref{KStarOverKVsData}, which is understood to be a consequence of the natural correlation between $N_{PI}$ and multiplicity. Also in this case, the suppression caused by the introduction of CR for large numbers of partonic interactions is observed to be of up to 7\% for the `More QCD-based' scheme, which indicates that events selected via $N_{PI}$ are equally affected by CR if compared to events selected via charged particle multiplicity, as done for Fig.~\ref{KStarOverKVsData}. 

It has to be noted, however, that a significant fraction of the charged kaons entering the denominator of the K*/K comes from resonance decays. 
In order to isolate the effect of CR on the prompt K*/K, it is interesting to remove all decay products from the calculation of this ratio, as also shown in Fig.~\ref{decayissue}, represented by the circle symbols. 
Once decays are removed, the overall K*/K ratios increase significantly but the overall trend of K* suppression when enabling CR is preserved. 
Because the K* also decays into charged kaons, the removal of resonance decay products from the ratios also increases the effect of CR to up to a 10\% decrease of the K*/K ratio at high values of $N_{PI}$.

We have also studied the effects of CR on yields as a function of transverse momentum, and have found that the suppression caused by CR for events with large $N_{PI}$ concentrates in the low $p_{T}$ region ($p_{T}$ $<$ 1.5 GeV/c), and affects both the yields of K* and K, but the suppression is more pronounced for the K*, leading to the decrease observed in the $p_{T}$-integrated K*/K ratio. 

\section{String hadronization study}

\begin{figure*}\begin{center}
\includegraphics[width=0.7\textwidth]{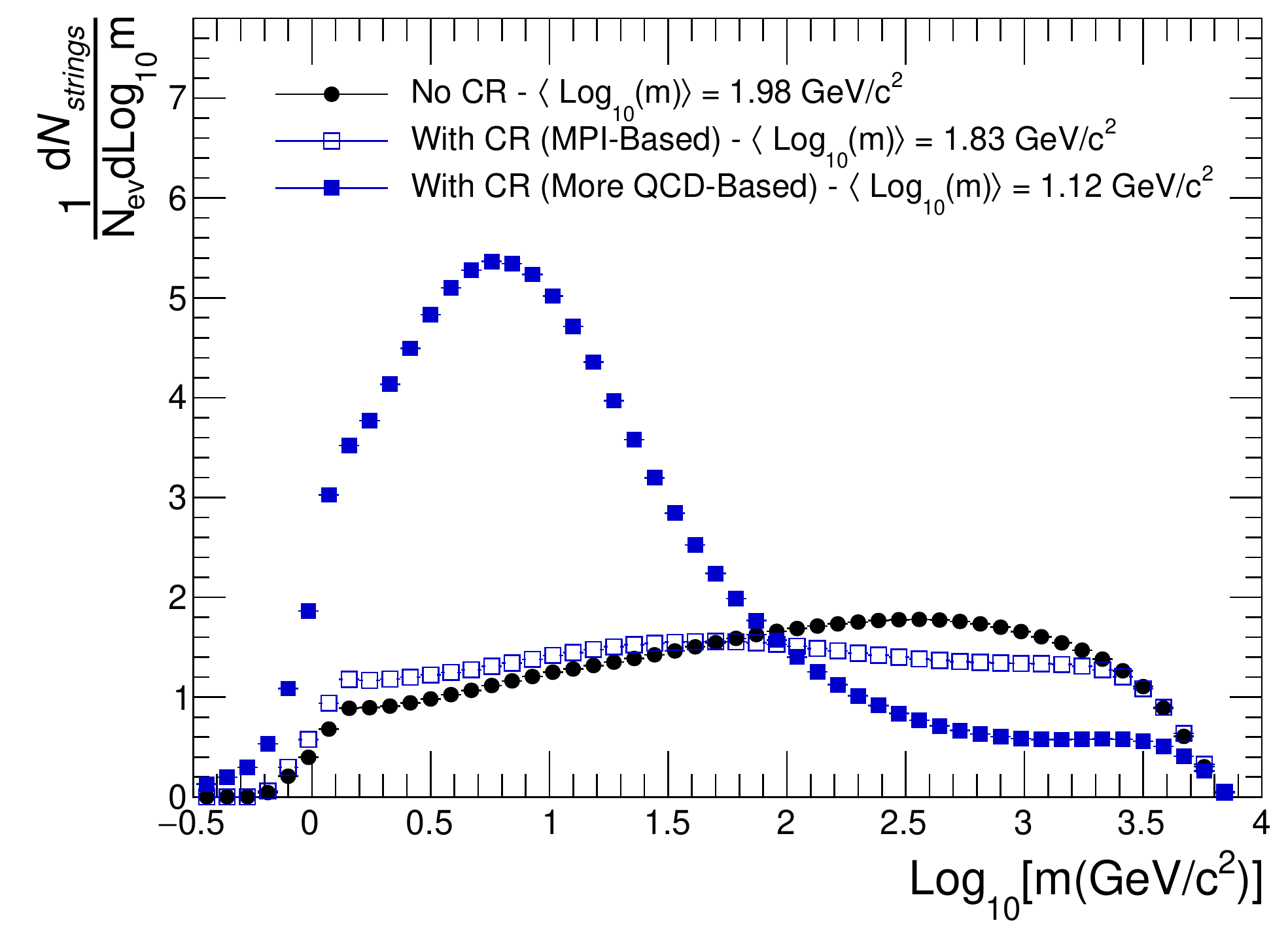}
\caption{\label{stringmassdist} String invariant mass distribution for inelastic events with 
at least one charged particle within $|\eta|<1.0$ generated
with the PYTHIA8 event generator with (blue symbols) and without (black symbols) CR. In both cases, the distributions were obtained without any selection 
on $N_{PI}$.}
\end{center}\end{figure*}

Considering that the suppression of K* mesons with respect to K is already apparent prior to all hadronic 
decays, it is clear that the effect must be related to string fragmentation. We studied the string invariant mass
distributions for simulated events with and without CR and as pointed out by \cite{ moreqcd}, we confirmed 
that enabling CR leads to a larger
population of lower mass strings, as can be seen in Fig.~\ref{stringmassdist}, with the most pronounced 
difference taking place for the `More QCD-based' CR scheme. This is to be expected considering 
that string length is directly proportional to the string mass in the PYTHIA8 event generator and that CR leads to 
shorter strings. 

\begin{figure*}\begin{center}
\includegraphics[width=0.7\textwidth]{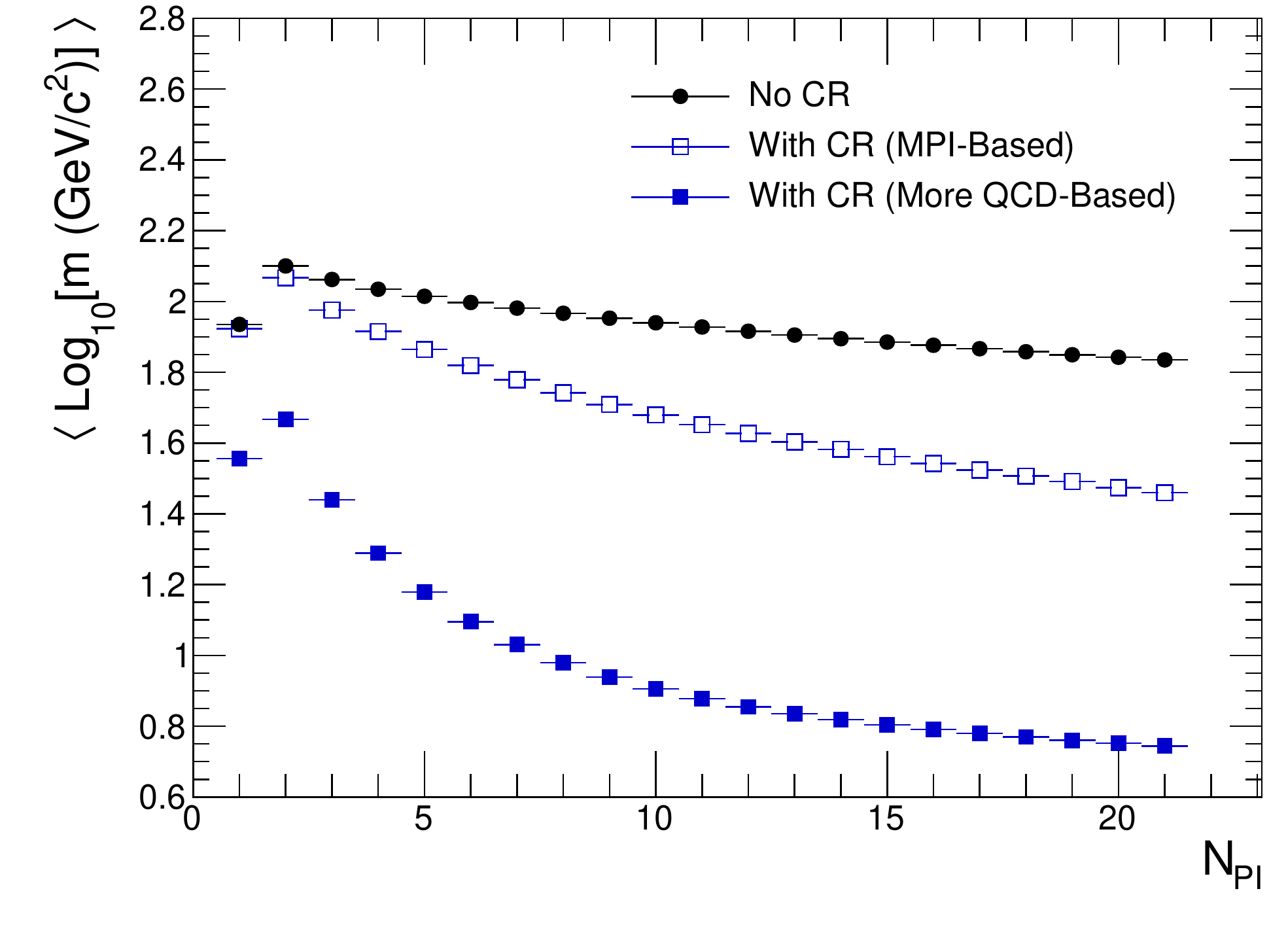}
\caption{\label{avermassvsNPI} Average string invariant mass as a function of number of partonic
interactions with (blue symbols) and without (black symbols) CR.}
\end{center}\end{figure*}

Because CR has a more pronounced effect for events with larger number of partonic interactions, the 
decrease in average string mass is also more apparent in this class of events, as can be seen in Fig.~\ref{avermassvsNPI}, which shows the average of the logarithm of the string invariant mass as a function 
of the number of partonic interactions. 
This suggests that any effect caused by the shortening of fragmenting strings will be more pronounced 
in high multiplicity pp events and will also be more pronounced in the `More QCD-based'- than in the 
`MPI-based' CR scheme. 

\begin{figure*}\begin{center}
\includegraphics[width=0.7\textwidth]{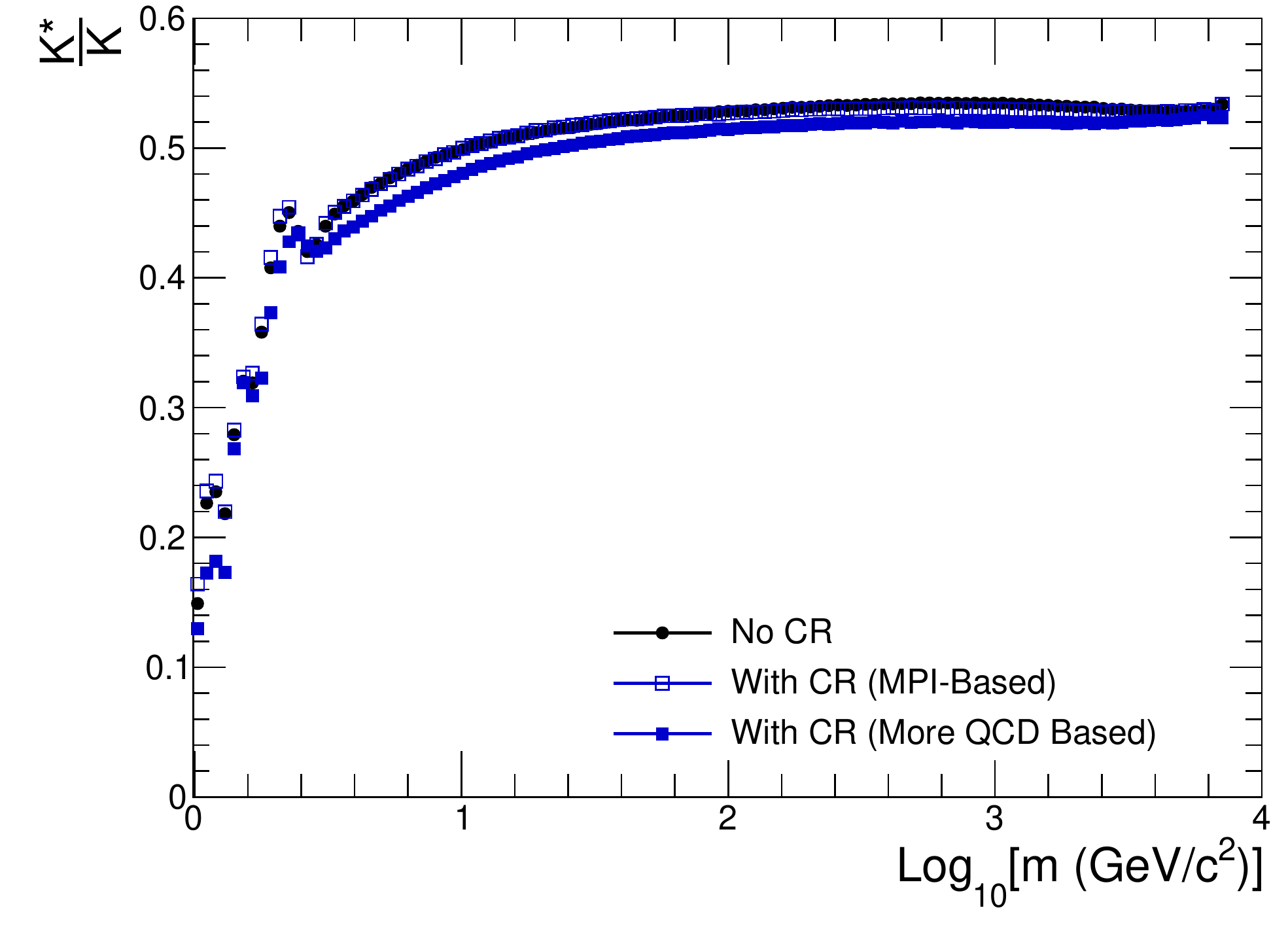}
\caption{\label{KStarOverKVsMass} Integrated K*/K ratio obtained when fragmenting strings of different
invariant masses with the PYTHIA8 event generator with (blue symbols) and without (black symbols) CR. No restriction on hadron rapidity was required for 
either K* or K.}
\end{center}\end{figure*}

In order to determine if lower mass strings lead to alterations in the resonant to non-resonant identified 
particle ratios, we calculated the K*/K ratio produced by fragmenting strings as a function of their invariant mass, 
as can be seen in Fig.~\ref{KStarOverKVsMass}. In order to avoid kinematic biases, no requirement on K* or K rapidity was 
applied when calculating these predictions.
The ratio is seen to be of approximately 0.54 for very massive strings 
but decreases to approximately 0.45 for low mass strings, with only little difference among the various CR schemes, which 
can be understood as a consequence
of the fact that lower mass strings are less likely to have enough energy while fragmenting to create a more
massive hadron, such as the K*, over a non-resonant charged kaon with the same quark content. 
The range in which the 
K*/K is seen to vary as a function of string mass also brackets the variation of this ratio when studied as 
a function of multiplicity, indicating that most of the suppression observed in PYTHIA is due 
to a shift in string masses. Furthermore, strings with lower masses also lead to particles with smaller $\langle p_{T} \rangle$, 
explaining why the suppression is observed to be strongest at low transverse momentum as mentioned
in the preceding section. 

\begin{figure*}\begin{center}
\includegraphics[width=0.49\textwidth]{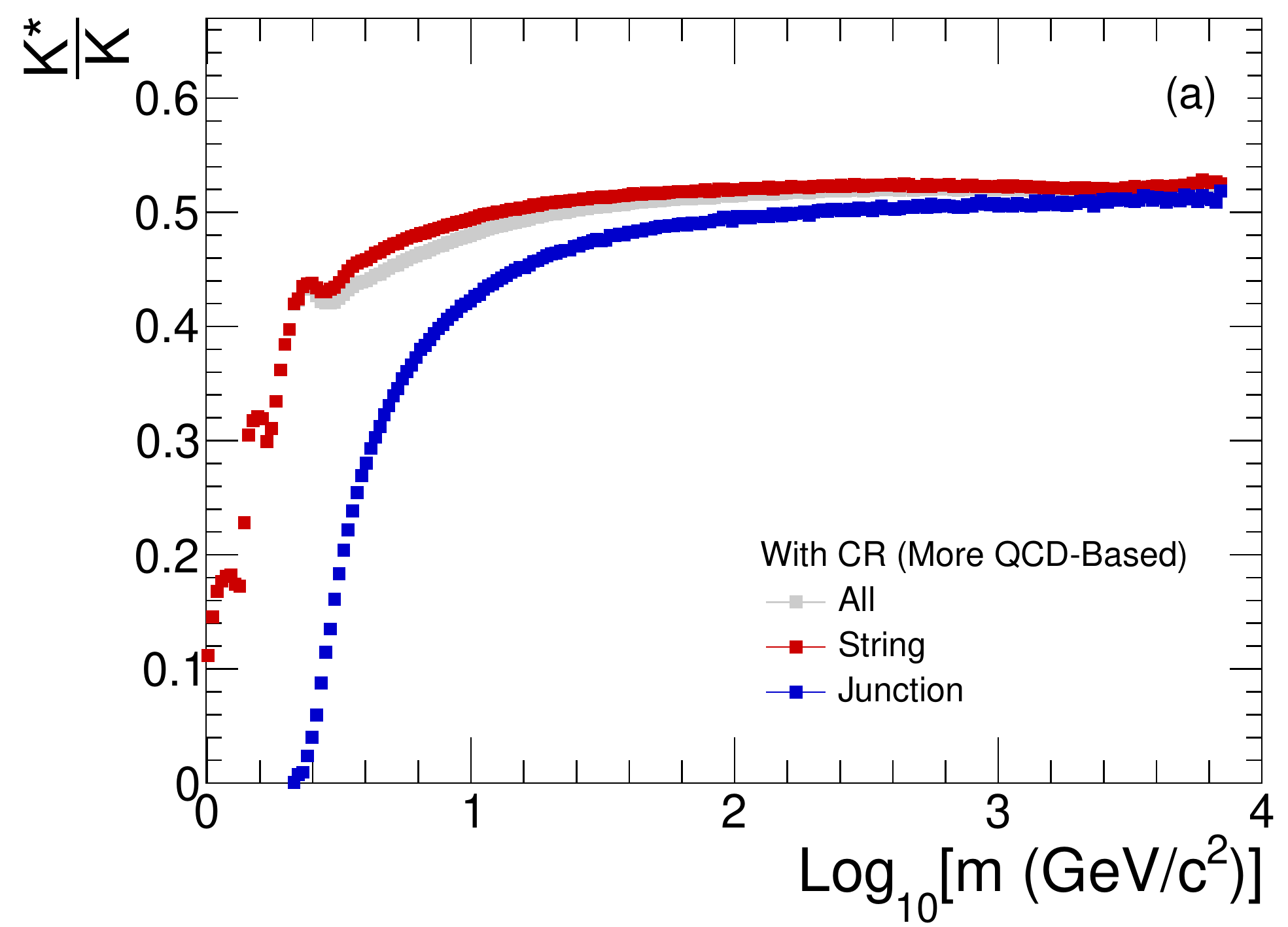}
\includegraphics[width=0.49\textwidth]{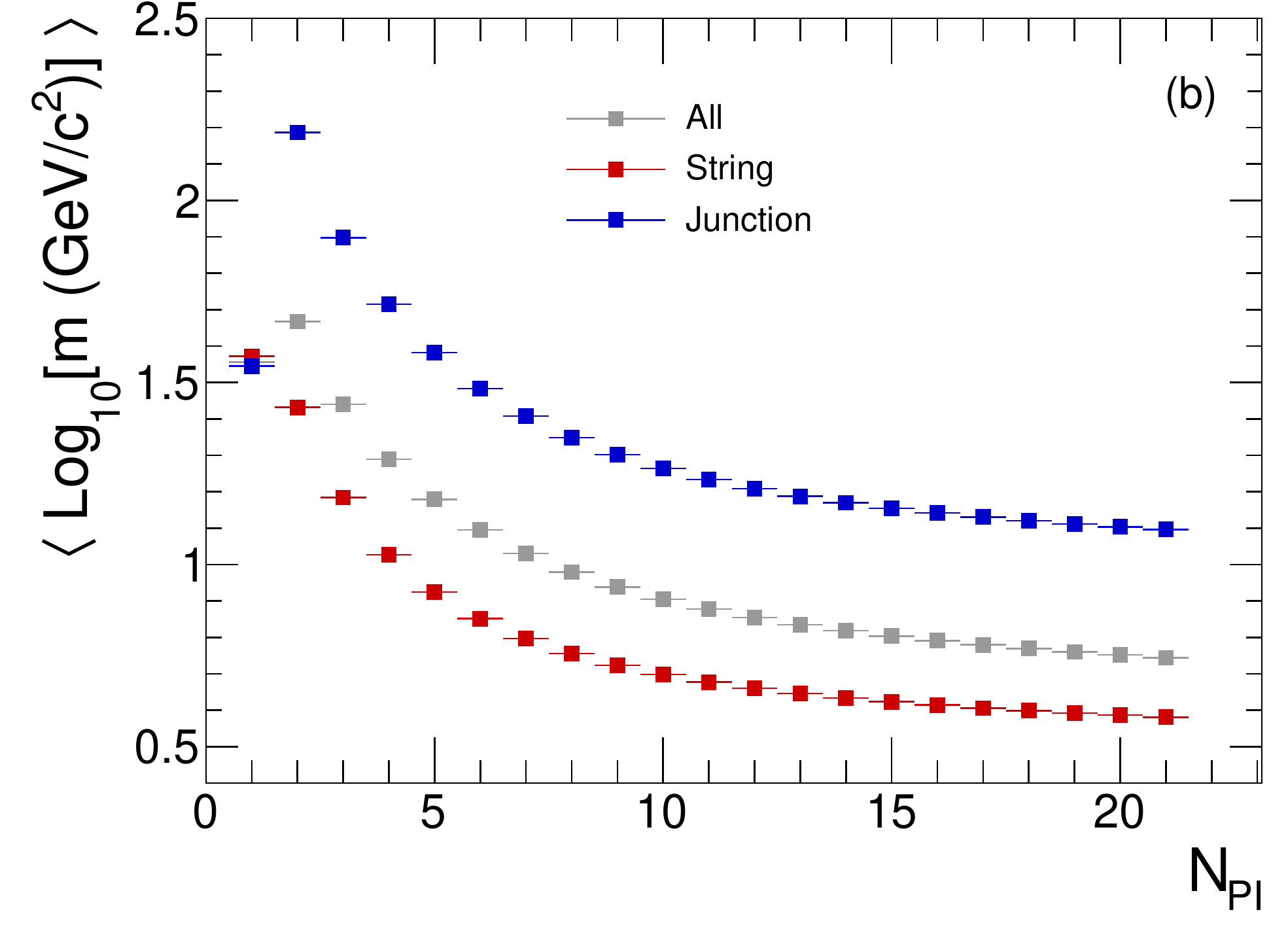}
\caption{\label{StringVsJunction} (left) Integrated K*/K ratio and (right) average string invariant mass as 
a function of number of partonic interactions for $q\overline{q}$ strings (red) and junctions (blue) in the `More QCD-based' CR scheme.}
\end{center}\end{figure*}

In addition to particle production from $q\overline{q}$ string fragmentation, PYTHIA is also 
able to fragment junctions into hadrons, a phenomenon that is responsible for up to 30\% of all charged particle production in the `More QCD-based' CR scheme in which junctions are most abundant. In 
order to study if junctions play a significant role
in the suppression, we have studied the K*/K ratio for $q\overline{q}$ strings and junctions separately, as can
be seen in Fig.~\ref{StringVsJunction}a. While in both cases the K*/K is seen to decrease with 
string mass, this decrease is more pronounced for junctions, which is a consequence of the fact that 
junctions favor baryons over mesons when fragmenting into higher mass hadrons. However, 
because $q\overline{q}$ 
strings are still responsible for over two thirds of charged particle production, the K*/K ratio obtained
for all color confining objects, shown in gray in Fig.~\ref{StringVsJunction}a, follows the trend of 
$q\overline{q}$ strings. Some structures are observed in the K*/K 
ratio at very low string masses, with the first of these structures appearing at $2.15$~GeV/$c$. 
These fluctuations in the ratio are due to the different treatment of very low mass objects employed 
by PYTHIA, in which 
such objects are called `mini-strings'. At progressively smaller masses, further structures in the ratio are 
to be expected, as string masses are of the same order of the hadrons 
into which they will fragment. Furthermore, as can be seen in Fig.~\ref{StringVsJunction}b, 
the average string mass is observed to become progressively smaller for larger $N_{PI}$ 
for both $q\overline{q}$ strings and junctions, indicating that in both
cases the decreasing string masses will be a central component of the suppression. Finally, the fact that
$q\overline{q}$ strings and junctions produce K* and K in somewhat different proportions for a given string
mass is also the reason for the small differences in the K*/K as a function of string mass for the various CR
schemes, seen in Fig.~\ref{KStarOverKVsMass}, as each scheme will have a different proportion of color 
confining objects. 

\begin{figure*}\begin{center}
\includegraphics[width=0.49\textwidth]{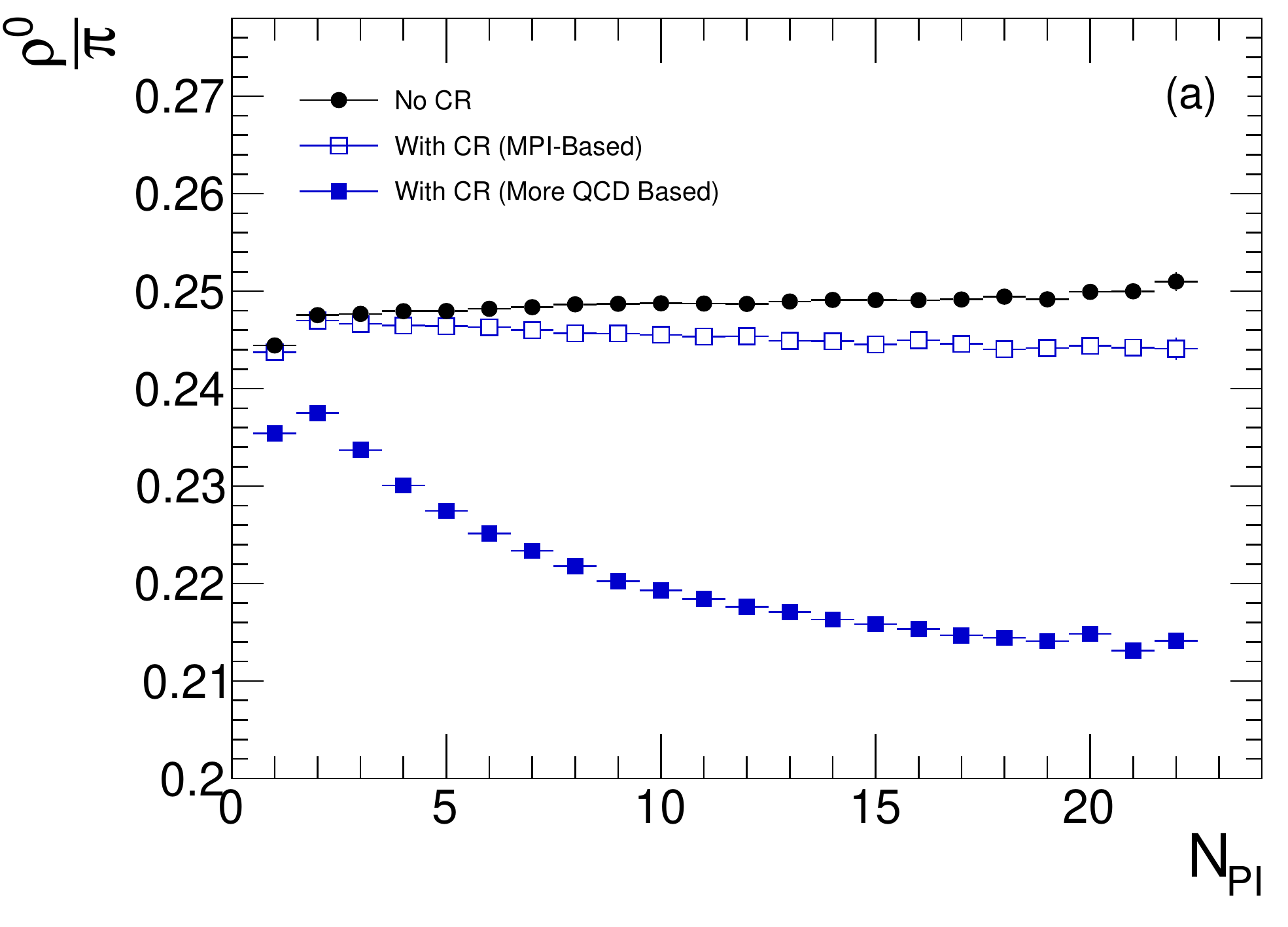}
\includegraphics[width=0.49\textwidth]{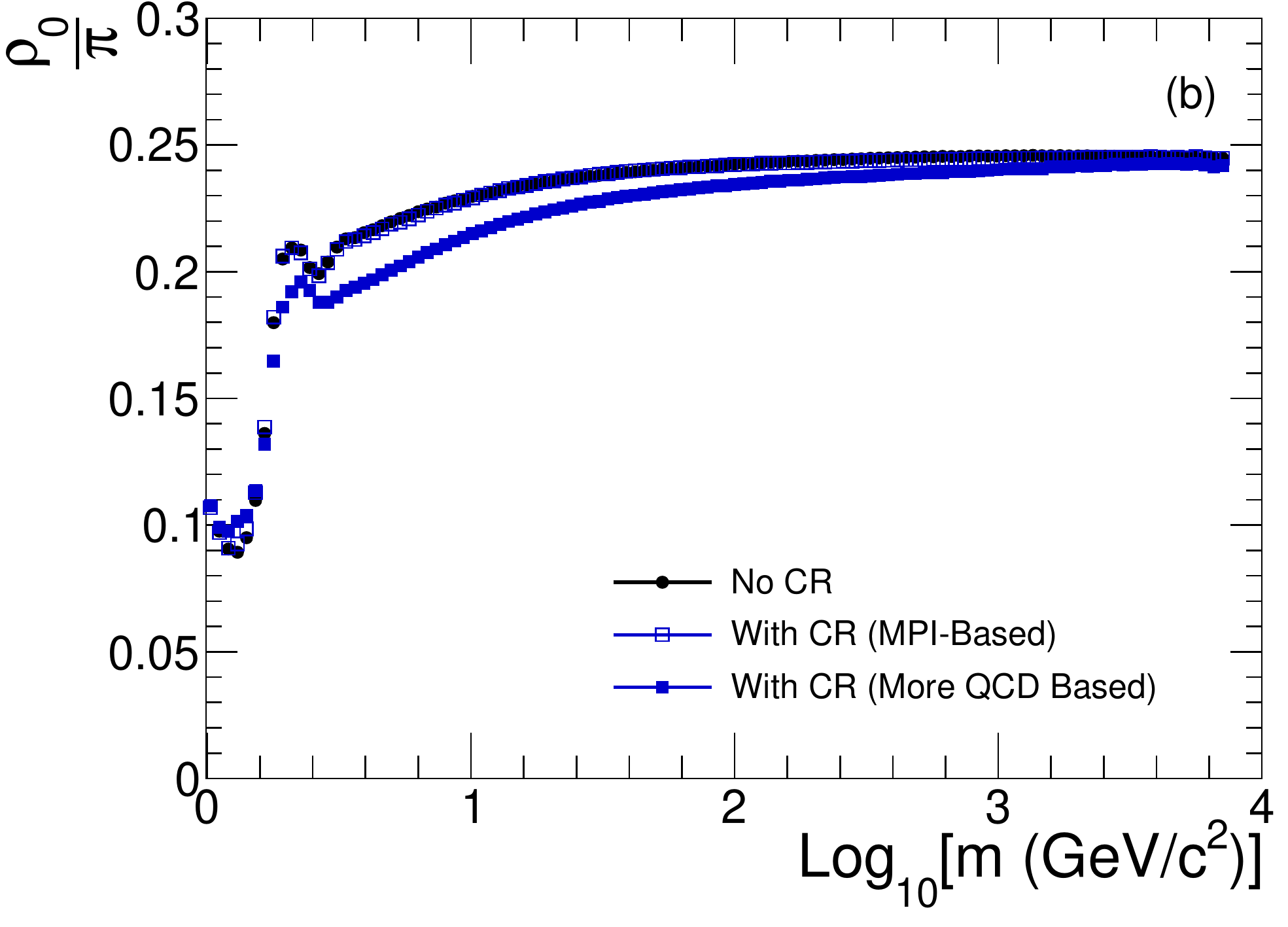}
\caption{\label{RhoZeroPlots} (left) Integrated $\rho_{0}/\pi$ ratio as a function of number of
partonic interactions and (right) as a function of fragmenting string mass with (blue symbols) and 
without (black symbols) CR.
All hadronic-level decays have been switched off explicitly for these predictions and no restriction on hadron rapidity was required
when calculating the $\rho^{0}/\pi$ ratio as a function of fragmenting string mass (right plot).} 
\end{center}\end{figure*}

We also studied the $\rho_{0}/\pi$ ratios, which again involve a meson resonance and a non-resonant state
with different masses but identical quark content. In this case, PYTHIA also predicts a suppression that 
gradually increases with $N_{PI}$, as can be seen in Fig.~\ref{RhoZeroPlots}a, and again the $\rho_{0}/\pi$ 
ratio is observed to decrease for strings with lower invariant masses, suggesting that, if measured, 
the $\rho_{0}$ yields should also be suppressed in high-multiplicity pp collisions. 

\begin{figure*}\begin{center}
\includegraphics[width=0.49\textwidth]{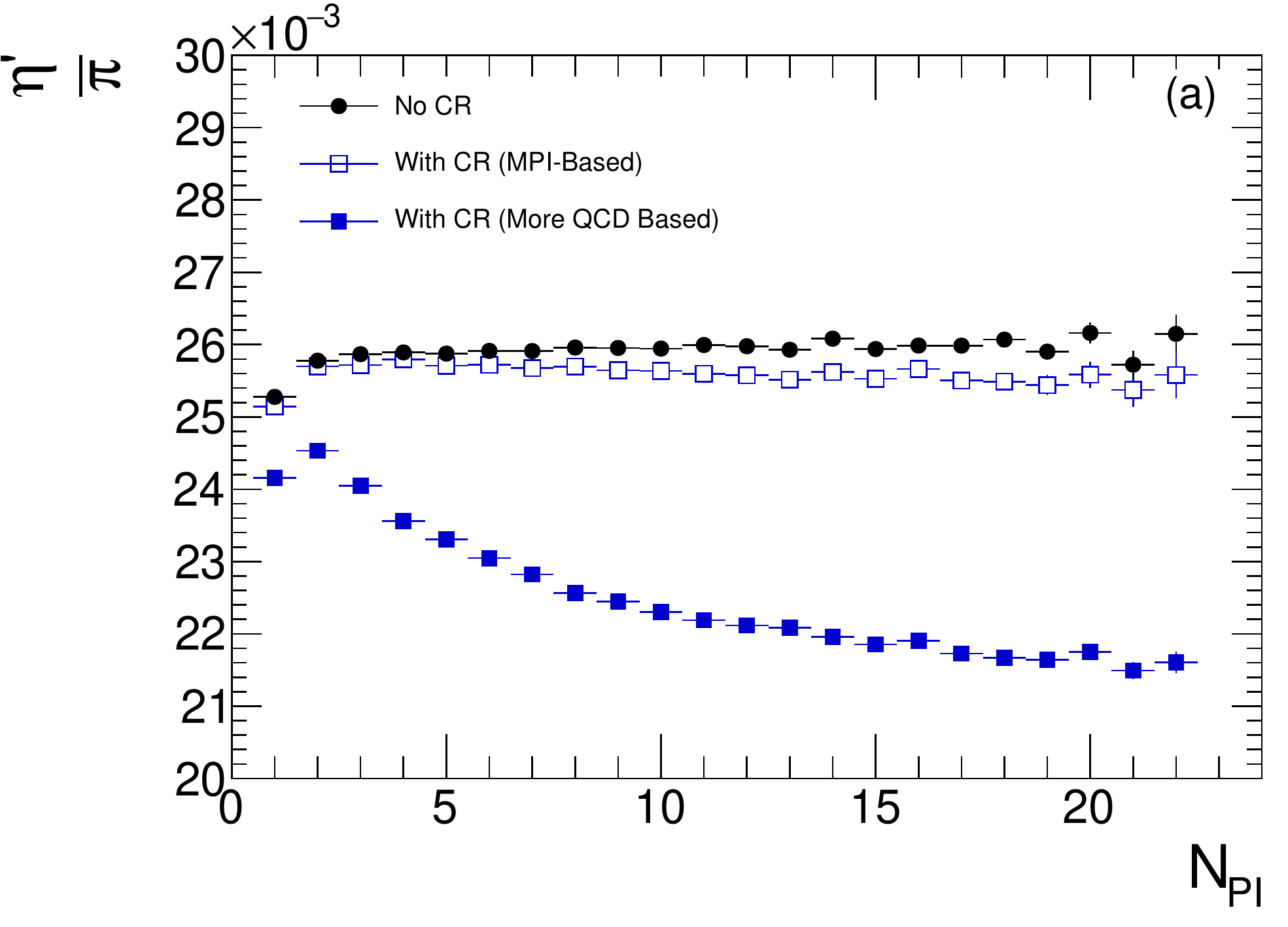}
\includegraphics[width=0.49\textwidth]{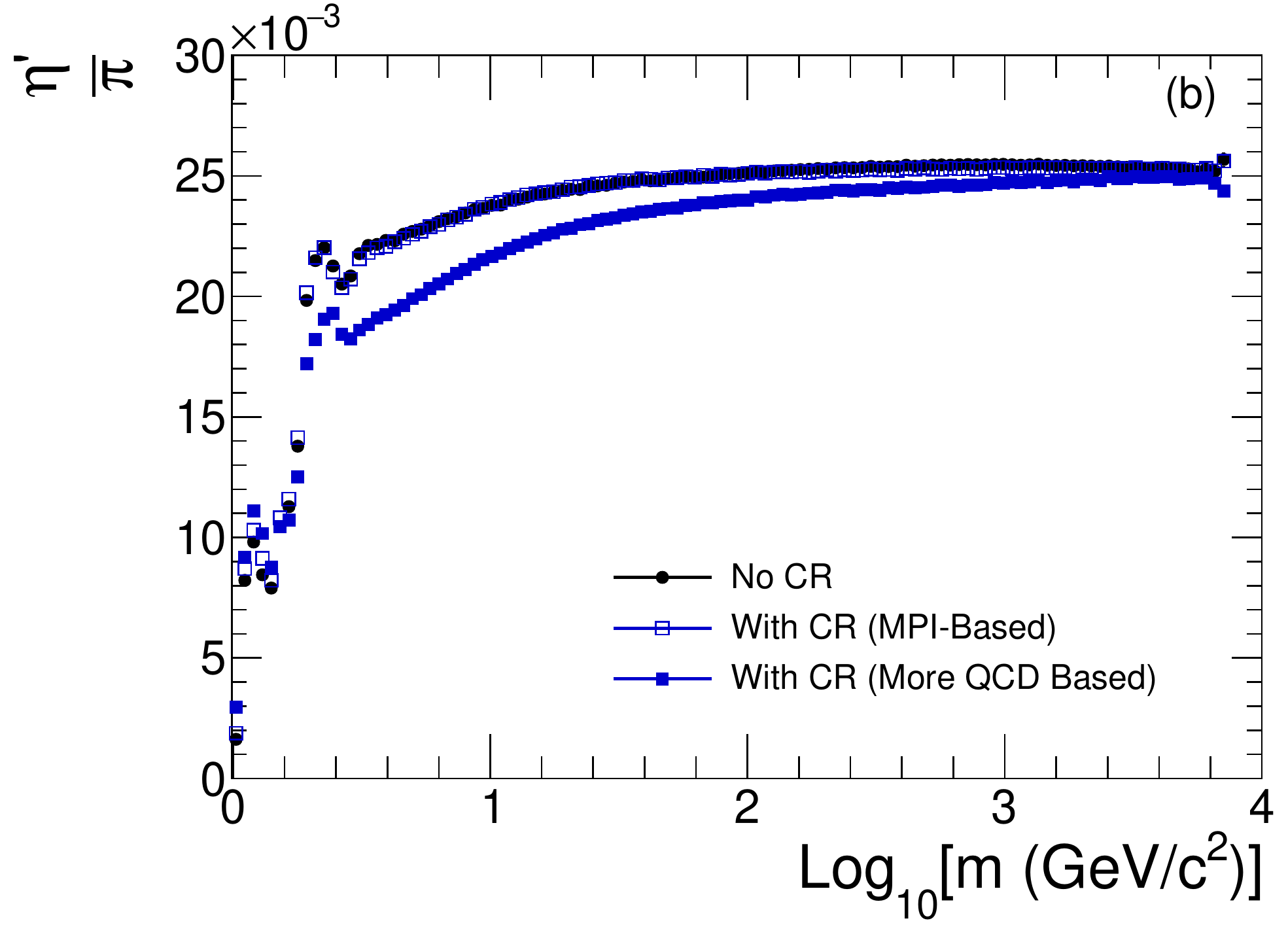}
\caption{\label{EtaPrimePlots} (left) Integrated $\eta^{\prime}/\pi$ ratio as a function of number of
partonic interactions and (right) as a function of fragmenting string mass with (blue symbols) and 
without (black symbols) CR.
All hadronic-level decays have been switched off explicitly for these predictions and no restriction on hadron rapidity was required
when calculating the $\eta^{\prime}/\pi$ ratio as a function of fragmenting string mass (right plot).} 
\end{center}\end{figure*}

In addition to a mass-related suppression, we also considered the possibility that 
a spin-related suppression could contribute to the decrease of K*/K. Such a mechanism 
could take place due to spin statistics and the fact that strings are spinless. In order to 
differentiate between a spin- and a mass-related effect, we also calculated the $\eta^{\prime}/\pi$ 
ratio, which is particularly interesting because it is a ratio of heavy to light mesons but in 
this case both are spin zero. As can be seen in Fig.~\ref{EtaPrimePlots}a, the $\eta^{\prime}/\pi$ ratio
is similarly suppressed as a function of number of partonic interactions, suggesting that 
the suppression is largely mass-related also for the $\eta^{\prime}$. 

A mechanism often used in complementarity with color reconnection is
the formation of `color ropes', combinations of more strings with combined
tension. This mechanism \cite{armesto, toporpop} is of
phenomenological interest since the increased string tension has been
shown to describe strangeness enhancement.
This scenario was also explored and we have found, using the DIPSY plug-in
for PYTHIA \cite{ropereference}, that also in this case resonances are 
suppressed in high-multiplicity pp collisions. However, it has to 
be noted that in this model string masses and lengths are no longer
proportional and therefore the decomposition of ratios as a function of 
string masses would not carry the same meaning as in the analysis discussed 
here. Physically this is reasonable since the formation of the ropes means the 
action of each string is characterized by both a string-specific length and a string-specific tension. 
We leave the interplay between these two factors to future work, considering the fact that 
the plug-in is not yet fully developed and tuned.

It has to be noted that the fact that the suppression is also present in this 
case is 
surprising at first sight: color ropes with increased local energy
density make it more likely for more massive final states, such as
strange hadrons, to form.  Hence, naively, they would go in the
direction opposite to the mechanism found here, favoring higher
mass states.
However, when one considers the physical basis for both color
reconnection and rope formation, the fact that they have the same
effect makes physical sense:
strings, in the underlying picture of the Lund model \cite{LundModel}, are a
`semi-classical' object locally minimising an action.
`Collective' effects, such as both reconnection and rope formation,
can be viewed as `tunneling' between the many local action minima
which arise in systems of many color charges.
Since the action is generally proportional to string tension, it is therefore
natural that rope formation, while increasing the
action locally, will tend to decrease it globally within the event.
And therefore its net effect on abundance of resonances relative to
lighter states with the same chemical composition will be the same.
This realization means the resonance suppression mechanism described
here is much more general than a simple `patch' in a Monte Carlo
phenomenological model, but should be general to all models having
such `semiclassical dynamics' in intermediate stages.  We shall
further discuss this universality in the final section.

\section{Conclusion}


Using the PYTHIA8 event generator, we checked the effect of MPI and CR on the production of resonance particles. 
We observed that CR suppresses meson resonance production in comparison to the non-resonance meson 
with the same quark content. 
Therefore, CR also serves as an alternate explanation for mesonic resonance suppression, which has already 
been measured by the ALICE collaboration in pp, p-Pb and Pb-Pb collisions but is usually explained only via 
rescattering and regeneration in a hadronic phase. In addition to the results shown here, we have also studied the effects of CR
on baryon resonance and non-resonance yields. In that case, other effects, such as the presence of junctions as in 
the `More QCD-based' CR scheme, will 
increase baryon production and thus offset any suppression. It has to be noted that, in any case, baryon production 
is significantly more model- and parameter-dependent than meson production and 
a comprehensive study regarding baryon production lies outside of the scope of this paper. 

The mesonic resonance suppression has been verified to be a consequence of the fact
that CR leads to shorter, less energetic strings which are less likely to fragment into resonances because 
these require more energy to be produced. Since CR is more pronounced for events with a larger number
 of partonic interactions, the intensity of this phenomenon is also multiplicity-dependent and is more pronounced
for high-multiplicity pp collisions. It is also strongest for models which predict shorter strings on the average, such as 
the `More QCD-based' CR scheme, in which junctions are allowed to form. Further study and tuning
of the interplay between junctions and $q\overline{q}$ strings might help in reproducing experimental 
observations and is left for future work.
It should also be noted that the suppression itself is a consequence 
of a physical mechanism that is not exclusive to any particular CR model, being present even if color 
ropes are present. 
Furthermore, cluster-based models such as HERWIG \cite{bahr-herwig} should give
qualitatively similar results as long as the clusters are allowed to fragment
and combine to minimize some rigorously defined free energy that is analogous to
string tension and length in string-based models. It is also reasonable to assume that CR effects will 
influence high density environments such as the ones created in heavy-ion collisions. Further 
studies would be needed to determine if the total suppression observed in ALICE across all systems 
can be explained as due to a combination of hadronization of shorter strings and hadronic re-scattering
or if only one of these effects is enough to adequately describe existing measurements. 

The authors acknowledge support from FAPESP grants 16/13803-2, 14/09167-8 and 17/05685-2. 
This work was also supported by U.S. Department of Energy Office of Science under contract number DE-SC0013391.

\end{document}